# How river rocks round: resolving the shape-size paradox

G. Domokos[1], D. J. Jerolmack[2*], A.Á. Sipos[1] and Á. Török[3]

[1]*Department of Mechanics, Materials and Structures, Budapest University of Technology and Economics, Műegyetem rkp 3, H-1111 Budapest, Hungary*

[2]*Department of Earth and Environmental Science, University of Pennsylvania, 240 S. 33rd St., Philadelphia, PA 19104, USA*

[3]*Department of Construction Materials and Engineering Geology, Budapest University of Technology and Economics, Műegyetem rkp 3, H-1111 Budapest, Hungary*

*\*corresponding author:* sediment@sas.upenn.edu*; (+1) 215-746-2823.*

**Abstract.** River-bed sediments display two universal downstream trends: fining, in which particle size decreases; and rounding, where pebble shapes evolve toward ellipsoids. Rounding is known to result from transport-induced abrasion; however many researchers argue that the contribution of abrasion to downstream fining is negligible. This presents a paradox: downstream shape change indicates substantial abrasion, while size change apparently rules it out. Here we use laboratory experiments and numerical modeling to show quantitatively that pebble abrasion is a curvature-driven flow problem. As a consequence, abrasion occurs in two well-separated phases: first, pebble edges rapidly round without any change in axis dimensions until the shape becomes entirely convex; and second, axis dimensions are then slowly reduced while the particle remains convex. Explicit study of pebble shape evolution helps resolve the shape-size paradox by reconciling discrepancies between laboratory and field studies, and enhances our ability to decipher the transport history of a river rock.

**Introduction**
Transport of pebbles in a stream causes them to collide and rub against one another and the stream bed, and the resulting abrasion produces the familiar smooth and rounded shape of river rocks. Pebble shape evolution due to abrasion has been a topic of study since Aristotle [1], yet there are few quantitative experiments and even fewer theoretical predictions. There are important consequences of the abrasion process: mass loss alters pebble mobility and hence can influence the form and evolution of a river profile [2,3]; abrasion produces sand and silt [3-6] that is deposited in downstream channels, floodplains and the ocean; and the degree of rounding observed in pebbles of fossilized stream beds is used to infer ancient river flow conditions [7]. Sternberg [8] was the first to report the now well-known result that pebble size decreases exponentially with distance downstream, a phenomenon he attributed to abrasion. Since that time, controversy has ensued regarding the importance of abrasion versus size-selective sorting in diminution of particle size [5].

The emerging consensus has been that abrasion rates reported from laboratory experiments [3-5,9-11] are too low to account for the downstream fining observed in natural rivers [12-14]; however, the few studies conducted with more energetic collisions – representative of steep river environments – reported significantly higher abrasion rates [3,11] . Other experiments have shown that size-selective sorting – in which small particles travel farther downstream than large particles – alone is sufficient to explain fining trends observed in the field [15,16]. Herein lies a paradox: there is clear evidence for significant mass loss from abrasion expressed in pebble shape, while pebble size trends are interpreted to suggest that mass loss from abrasion is negligible. However, most field studies are not directly comparable to laboratory results; the former typically measure the length of only one pebble axis, while the latter report mass loss [5]. As pointed out by several researchers [4-5], rounding of a cube to an inscribed sphere would reduce mass to π/6 of its original value while producing no change in measured axis lengths. These authors concluded that the importance of abrasion may be significantly underestimated by field studies. Clearly, shape must be explicitly considered when assessing the contribution of abrasion to downstream fining of sediments [17]. Several recent experiments that examined shape evolution under abrasion [18,19] provided qualitative confirmation of geometric models [20-22], which predict that regions of high curvature are preferentially eroded. Building on this work, we present the first quantitative test of the curvature-driven abrasion model originally proposed by Firey [20], using laboratory experiments and a discrete chopping model. Experiments show unequivocally that abrasion occurs in two phases depending on particle shape. This two-phase behavior emerges spontaneously from the both the continuous-Firey and discrete-chopping models.

**Curvature-driven abrasion: background and theory**
For simplicity we focus in this study on the limiting case of a pebble colliding with a flat plane (or alternatively, striking the surface of an infinitely large abrader), which could approximate the situation of a cobble on a smooth bedrock river bed. To further clarify the controls of shape on abrasion, our experiments and models begin with cuboid particles. Abrasion is assumed to be isotropic; that is, all collision directions are equally probable. The geometric control of curvature in this situation is intuitive: areas having positive curvature protrude from the pebble and thus will be abraded (Fig. 1a). On the other hand, areas of non-positive curvature will not impact the plane and therefore won't abrade. We may qualitatively anticipate two phases of abrasion, where high-curvature regions are first removed and then the rounded pebble slowly reduces its size. This intuitive picture has been confirmed qualitatively by experiment [4,5,10,18].

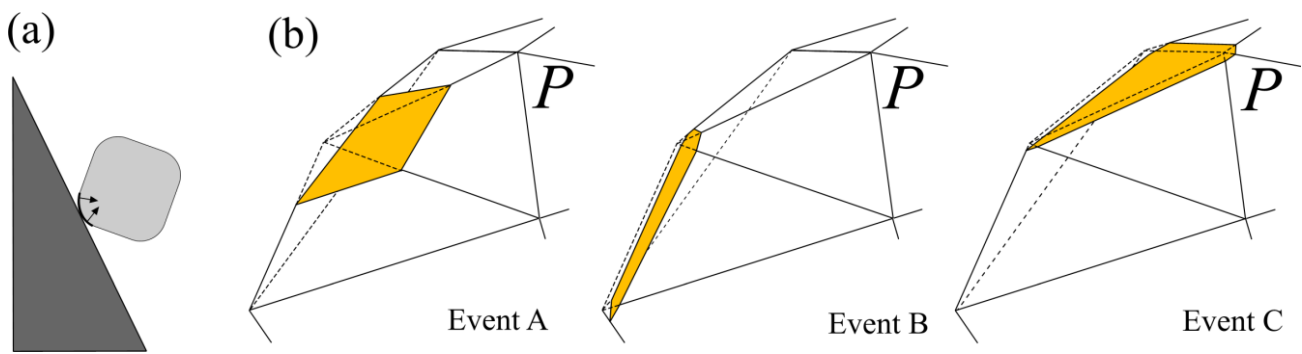

**Figure 1.** Definition sketch. (a) 2D schematic of the physical situation studied, showing an abrading cuboid colliding with a flat plane. Zone of positive curvature on the colliding corner is highlighted with arrows indicating surface-normal abrasion. (b) Three scenarios of the chopping model: Vertex chopping (Event A) corresponding to Gaussian-curvature-driven abrasion, edge chopping (Event B) corresponding to Mean-Curvature-driven abrasion, and face chopping (Event C) corresponding to uniform (Eikonal) abrasion.

In order to mathematically model abrasion and compare to experiments, a precise and parsimonious description of pebble shape is required. Based on the concept of the convex hull – the smallest convex body containing the original (non convex) shape – we introduce the surface convexity index $\beta = A_C/A_H$, $0 \leq \beta \leq 1$, where $A_C$ is the area of the convex regions and $A_H$ is the total area of the convex hull. (The previously-derived volume convexity index [23] is constant for our situation, since abrading cuboids always exhibit non-negative curvature; it is therefore not considered here). We set $a \geq b \geq c$ as the lengths of the principal axes of the pebble (Fig. 2), where the ratios of the major axes $y_1 = c/a$ and $y_2 = b/a$ are useful additional shape parameters [24]. Shape parameters that provide further information, and allow comparison to previous experimental and theoretical work, are: the Wadell sphericity, $r$ [25]; and the exponent $n$ corresponding to the best-fitting superellipsoid, given by $(x/a)^n + (y/b)^n + (z/c)^n = 1$. The anticipated two phases of abrasion may now be formally defined: (I) the polyhedral (non-convex) phase with β < 1 and constant $y_1$ and $y_2$; and (II) the smooth (convex) phase with β = 1 and non-constant axis ratios (Fig. 2).

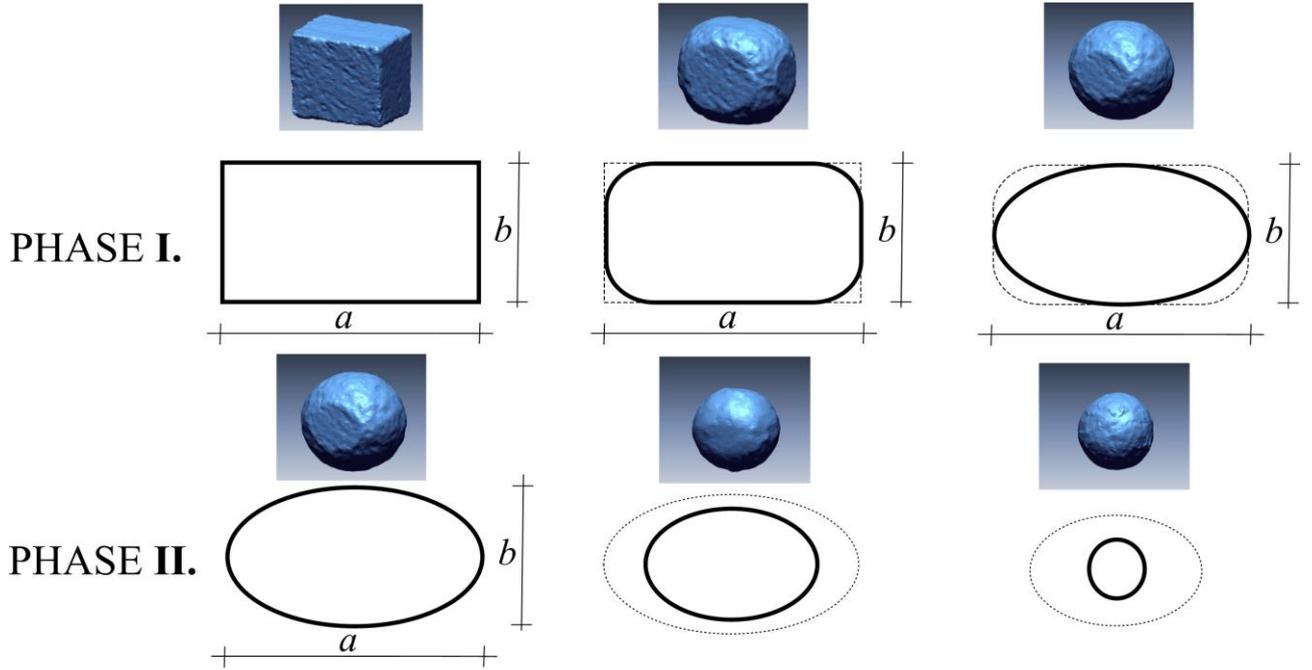

**Figure 2.** Two-phase abrasion illustration. The 2D schematic shows two well-separated phases emerging spontaneously from Gaussian-curvature-driven abrasion: In Phase I edges abrade but axis ratios remain constant; in Phase II, axis ratios evolve towards the sphere. Accompanying perspective images are topographic laser scans that illustrate the two phases in 3D; they were performed for a separate experiment with a smaller cuboid having similar axis ratios.

Firey [20] derived a geometric partial differential equation (PDE) to model shape evolution of a convex particle abraded by repeated collisions with an infinitely large abrader (i.e., a flat plane). In this model, local abrasion occurs in the direction normal to the surface at a speed $v$ that is proportional to the Gaussian curvature $K$:

$v = gK.$  (1)

Bloore [21] generalized this model to accommodate finite-size abraders; in three dimensions (3D) the evolution equation becomes:

$v = 1 + 2fH + gK,$  (2)

where $H$ is the mean curvature, and $f$ and $g$ are the integrated mean curvature and surface area of the abrader, respectively [26]. Note that in the limit of a very large abrader, **Eq. 2** reduces to **Eq. 1** and the limiting geometry is a sphere. In the limit of very small abraders, Eq.2 reduces to parallel flow with constant speed; shapes move away from the sphere and the limiting geometries have flat faces and sharp edges [30]. Friction from sliding may also contribute to surface-parallel erosion, preventing abrading pebbles from converging to a spherical shape. For the general case ($f,g>0$) normal- and parallel-flow effects compete; however, surface-normal abrasion dominates for generic collision processes [21,29]. Our experiments approximate a series of collisions with a very large object (drum), thus we expect that **Eq. 1** is adequate to describe pebble abrasion. However, we implement a numerical solution to the general **Eq. 2** to test this assumption and the ability of this curvature-driven model to reproduce the two-phase abrasion observed in experiments.

**Results**

We performed a series of four laboratory experiments in which single cuboids composed of oolitic limestone (initial size [mm] $a_0$ = 70.8 ±0.8, $b_0$ = 60.7 ±0.7, $c_0$ = 50.6 ±1.2; initial volume $V_0$ = 217,456 ±10000 mm$^3$) were abraded in a 1-m diameter rotating drum, to simulate collision of a pebble with an infinitely large abrader. A paddle in the drum lifted and dropped the particles once per rotation, thus preventing friction-induced abrasion from sliding. At specified rotation intervals we imaged each face of the pebble, and measured the three principal axis lengths plus the mass (Figure 3; see Methods). Experiments produced identical exponential declines in pebble mass with time (number of rotations) (Fig. 4f), consistent with expectations from previous experiments [3-5,10] that abrasion rate is proportional to kinetic energy of impact. To facilitate direct comparison of experimental results to geometric modeling, we assessed shape evolution as a function of pebble volume. A striking result is the clear emergence of the anticipated Phases I and II of abrasion (Fig. 4). This is most clearly expressed in the axis ratios (Fig. 4a, b), which were constant over the interval $V_0 \geq V > 140,000$ mm$^3$ (Phase I) but grew toward $y_1 = y_2 \to 1$ as volume was further reduced (Phase II). Convexity increased over the same interval indicating rounding; however it became constant ($\beta \approx 1$) for $V < 140,000$ mm$^3$ (Fig 4c). We also observed that the evolution of Wadell sphericity, $r$, tracked $\beta$ (Fig. 4d), whereas $n$ rapidly dropped toward $n \to 2$ (Phase I) and then remained at $n$ = constant $\approx 2$ (Phase II), the latter corresponding to ellipsoidal shapes (Fig. 4e). These data provide a benchmark for testing the geometric abrasion models.

Two complementary modeling approaches were undertaken to examine 3D pebble abrasion under conditions simulating the laboratory experiments. First, we modeled surface evolution of a cuboid by numerically integrating the PDE **Eq. 2** using a standard *level-set method* [27,28] (see Methods), with coefficients chosen to match the experimental data. The second approach involved implementing a stochastic, discrete-event *chopping model* introduced in [22]. For each collision, a prescribed volume (selected from a lognormal distribution) is removed from the pebble by randomly picking a collision direction and intersecting the pebble surface with a plane. Three types of events can occur (Fig. 1): (A) a vertex is chopped with an arbitrary plane and a polyhedron gains a new facet, with probability $p$ (similar to Gaussian curvature flow); (B) an edge is chopped with a plane parallel to that edge and a polyhedron gains a new facet, with probability $q$ (like mean curvature flow); and (C) a facet is shifted inward with probability $1-p-q$ (realizing uniform flow; Fig. 1b; see Methods). A higher Gaussian curvature ($p$), for example, would lead to a faster convergence to a sphere. The average surface abrasion rate $w$ over many collisions in the chopping model exhibits behavior convergent with **Eq. 2**:

$w = (1-p-q) + qH + pK.$ (3)

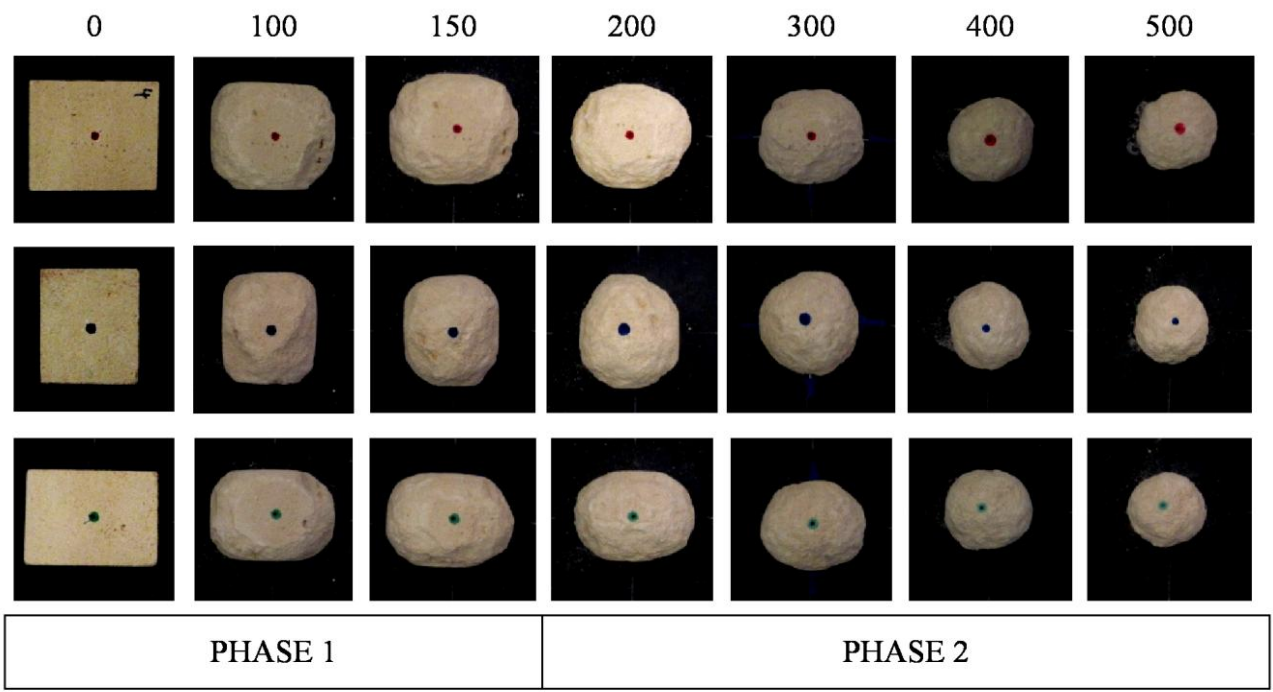

**Figure 3**. Experimental images of abrasion. Three rows correspond to three orthogonal views of the specimen, and columns show time evolution in terms of number of drum rotations. Separation of Phases I and II can be observed by visual inspection.

Two-phase abrasion emerges spontaneously from both the level-set and chopping models (Fig. 4), with the evolution of all shape parameters in good agreement with experiments. The level-set method produces a smooth trend, indicative of its idealized continuous representation of collisions, while the chopping model exhibits stochastic fluctuations similar to the experiments. We found that both models achieved an optimal fit to the data with pure Gaussian curvature flow ($p = 1$ and $q = 0$ for **Eq. 3**); i.e., Firey's model (**Eq. 1**). Phase I shows a sharp increase in convexity $\beta$ and constant axis ratios, while Phase II consists of $\beta = constant \approx 1$ and $y_1$ and $y_2$ increasing linearly with decreasing volume. The transition from Phase I to II in the models occurs for the same volume as observed in experiments. We also observed a sharp transition in the evolution of the exponent $n$ of the fitted superellipsoid: Phase I showed a fast drop of $n$ while Phase II exhibited almost constant $n \approx 2$ (Fig. 4).

Results for a cuboid have been presented for simplicity; however, the approach may be generalized to arbitrary shapes. Doing so reveals that the phenomenon of two-phase abrasion is robust. As an example, we use the chopping model [22] to simulate the evolution of a tetrahedron subject to abrasion from a Gaussian flow. As with the cuboid, convexity $\beta$ increases with decreasing pebble volume and, at $\beta = constant \approx 1$, the pebble reaches Phase II. One difference, however, is that the transition is less abrupt compared to cuboids (Fig. 5). For all shapes, the transition in phases coincides with the complete removal of the original facets from the polyhedron. This same phenomenon was also observed in Kuenen's experiments [10] of a cuboid rolling over a fixed pebbly bottom, driven by a water current: *"Up to the very last moment before a cube is rounded to a spherical shape, the untouched original faces can be recognized, and the diameter of the sphere is equal to the edge of the original cube to within a few tenths of a millimeter."* Thus, it appears that two-phase abrasion occurs even when the assumption of a flat plane is relaxed. Theory predicts that, even in the case of mutually-abrading and co-evolving pebbles, the curvature terms in **Eq. 2** will dominate the PDE [29,30]. Tumbling mill experiments with multi-pebble collisions [4] provide qualitative support for two-phase abrasion.

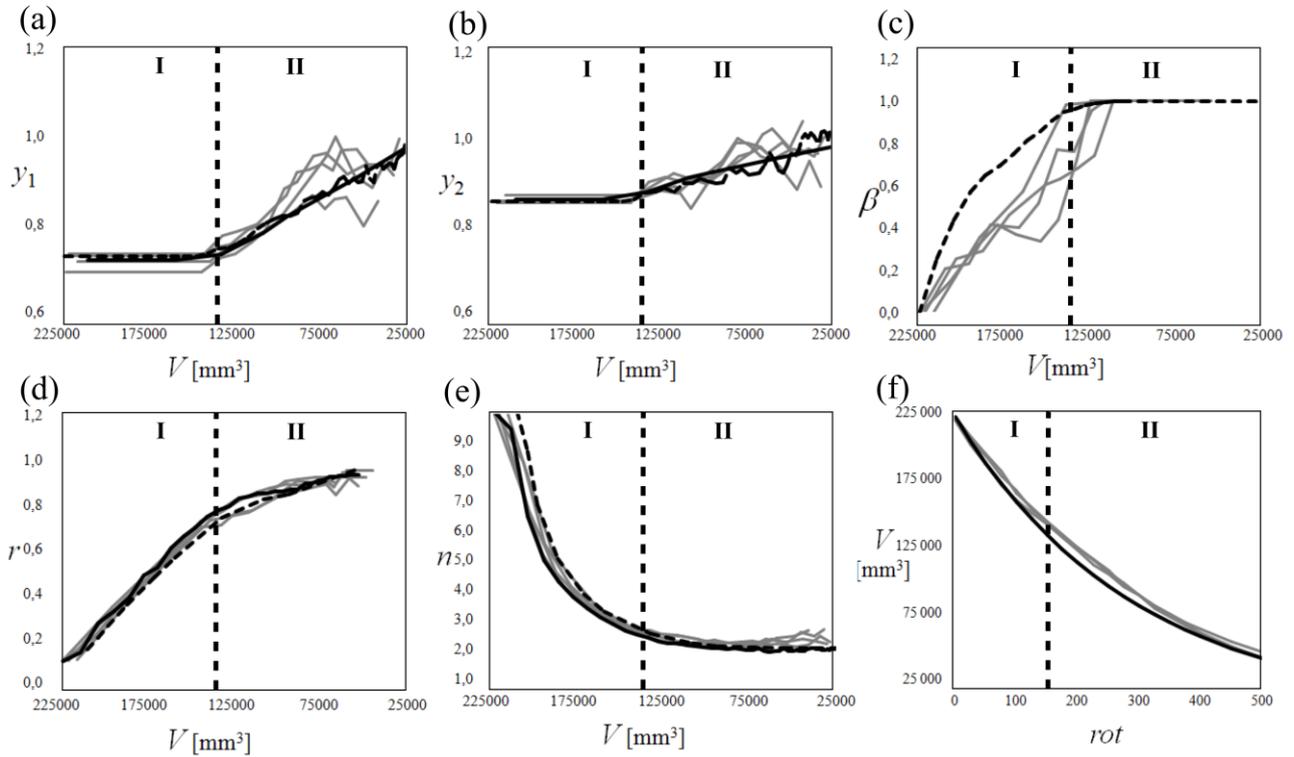

**Figure 4.** Comparison of experimental and numerical results. (a-e) Evolution of shape parameters versus volume, $V$. Shown are: axis ratios (a) $y_1$ and (b) $y_2$, (c) convexity index $\beta$, (d) Wadell sphericity, $r$, and (e) superellipsoid exponent, $n$. (f) Evolution of $V$ versus the rotation number (*rot*), a proxy for time. Gray line: experimental data. Black solid line: level-set method approximation of the PDE (**Eq. 2**). Dashed line: chopping model approximation (**Eq. 3**). Best fit coefficients correspond to pure Gaussian flow. Note abrupt change for all shape parameters (a-e) at transition from Phase I to Phase II, shown with vertical dashed line. Pebble volume exhibits no abrupt change through time (f); the fitted exponential trend is identical in Phase I and Phase II. Data used to generate this figure are contained in Supplementary Information File S1.

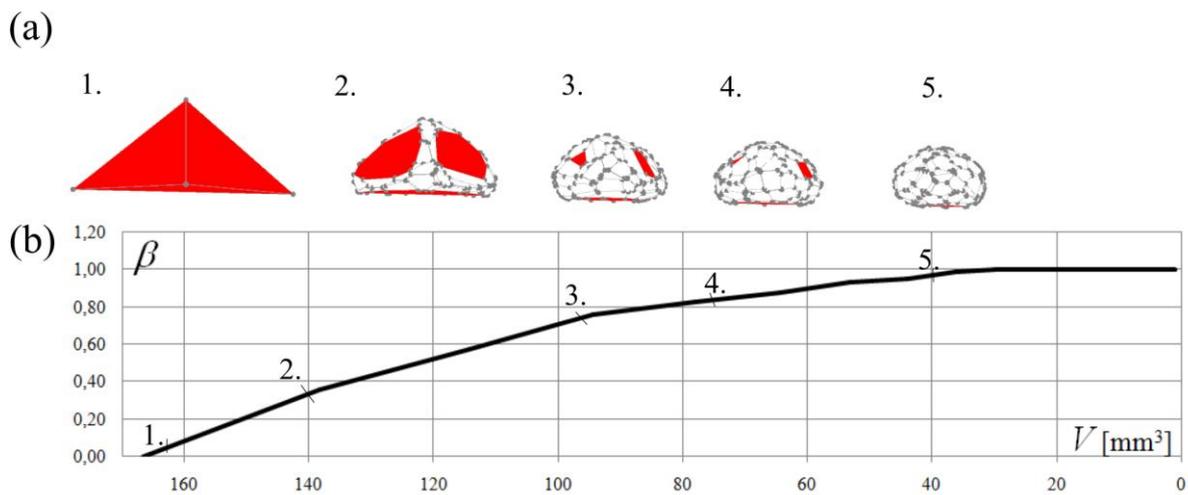

**Figure 5.** Shape evolution of a tetrahedron under the purely Gaussian chopping model in **Eq. 3** with $p = 1$. (a) Initial facets are shown in red; the transition to Phase II abrasion occurs when these facets have been entirely removed. (b) As with the cuboid, surface convexity $\beta$ increases during Phase I and stabilizes at $\beta \approx 1$ in Phase II.

**Discussion and conclusions**

A geometric model that prescribes abrasion rate simply as a function of local curvature (**Eq. 1**) shows that any initially-polyhedral particle will exhibit two phases of abrasion. Remarkably, the simple Gaussian flow description not only correctly predicts two distinct, well-separated phases of pebble shape evolution, but is also sufficient to quantitatively reproduce the shape evolution of a real abrading pebble. The differences in the evolution of β are due to the fact that, in our code, the abraded volume does not depend on the local geometry of the polyhedron, see **Methods.** The governing PDE (**Eq. 1**), obtained as the result of averaging over discrete collisions, is of generalized parabolic type (its linearized version is parabolic), and its qualitative behavior is perhaps best understood via the well-known heat equation [31,32]. In the analytical heat equation, heat from a discrete source propagates to the full domain at infinite speed, a phenomenon known as "instant smoothing". In our model, Gaussian curvature is analogous to heat. While instant action is obviously an unphysical artifact of the linearized PDE, it certainly *signals* a short but intensive burst in shape evolution – both in the original physical process and in its direct, discrete-event based simulations – and this burst corresponds to Phase I in our model. The mathematical essence of this phenomenon – in the context of the fully nonlinear PDE **Eq.1** – was first described by Hamilton [33], and it would be of prime interest for future research to compare the timescale of Phase I predicted by **Eq. 1** to the timescale predicted by the discrete collisional model and the experiments.

We may make some inference about the duration of Phase I from the physical system. While axis ratios and shape parameters show a marked jump at the phase boundary, volume diminution rate is not sensitive to the phase transition and can be well approximated by a single exponential curve (Fig. 4). "Sternberg's Law" [8], which predicts an exponential decay of pebble size with downstream distance along a river ($x$), is typically interpreted in terms of particle diameter; i.e., $a(x) = a_0 e^{-\alpha x}$. According to our findings pebble diameter is constant in Phase I, so this law is only valid in Phase II. Since pebble shape changes rapidly in Phase I, one would expect that the volumetric version of Sternberg's Law – i.e., $V(x) = V_0 e^{-3\alpha x}$ – is also only valid in Phase II. However, both our computations and experiments indicate that there is only a very small change in the exponent at the phase transition. This leads to an interesting generalization of Sternberg's Law: Volume evolution – but not diameter – may be approximated by a single exponential function throughout the entire abrasion process. Based on this observation we may estimate the length of Phase I in fluvial environments in terms of river length $x$ [km], using existing data on size diminution in Phase II. According to [13], abrasion-dominated rivers typically exhibit $\alpha < 0.03$; in terms of volume loss, Sternberg's Law is then $V(x) = V_0 e^{-3\alpha x}$. Based on our experiments and computations, we can constrain the minimum volume loss $\Delta V$ in Phase I as $\Delta V > 0.1 V_0$; the precise volume ratio depends on initial pebble shape. Using this value and an upper estimate of $\alpha = 0.03$ yields a minimal length for Phase I abrasion along the river, $x \sim 1$ km. This is in fair agreement with the few quantitative field and experimental studies of shape, which report that rapid rounding of pebble edges occurs within the "first few kilometers" of a stream [4,10,34]. Converting this distance to a timescale would require detailed knowledge of pebble transport and burial statistics in a river, which is beyond the scope of the present work [see 2]. We can however identify qualitative effects that may prolong Phase I compared to our single-particle, friction-free drum experiments: (a) Friction-dominated abrasion, in which flat areas are subject to sliding friction, preserve their flatness [29]; (b) small abraders cause the first (constant) term in **Eq. 2** to dominate, causing flattening of faces in a manner similar to friction; and (c) *collective abrasion,* where the coefficients $f$ and $g$ will co-evolve with the abraded particle [29,35], and the constant term will initially play a key role [36]. We also note that, in nature, even well-rounded pebbles are seldom spherical; non-spherical limiting shapes are predicted in models that combine collisional and frictional abrasion [29,35]. Finally, it should be clear that **Eq. 2** does not model fragmentation or crushing of pebbles.

This occurs most frequently in the energetic upper reaches of rivers, and has an effect opposite of abrasional smoothing [3, 11].

An important result from our work is that effective particle "size", as typically measured by axis lengths, is constant during Phase I abrasion – even though up to half of pebble mass is lost in this Phase – as the shape evolves toward that of an inscribed ellipsoid. In our experiments, Phase II abrasion is driving the particles towards the sphere; in a more general setting, under mutual abrasion of particles also subject to friction and size-selective transport, Phase II abrasion may result in the emergence of dominant axis ratios corresponding to non-spherical, ellipsoidal shapes [29, 36]. In nature, the rocks supplied from valley walls to a stream are typically very angular. Because of the common assumption that rapid rounding occurs in the first few kilometers of downstream transport [4,10], researchers have selectively neglected this effect in models and experiments [2,3]. In addition, virtually all field studies measure pebble diameters – rather than masses – to infer downstream diminution rates [5]. Both situations implicitly assume that Phase II abrasion alone is operative. Our results suggest that one may explicitly delineate where the transition from Phase I to II occurs in a river profile (or a laboratory experiment), by determining where (when) the average pebble surface convexity achieves a constant value close to 1. Axis ratio evolution in Phase II is predicted by the discretized version of **Eq. 2**, the so-called *box equations* [29], which allow determination of whether or not abrasion contributes to downstream fining for Phase II.

The extent to which these experimental and modeling results may be extended to the field is uncertain, considering the complexity of natural sediment transport and the heterogeneity of pebble material properties. Our findings are most relevant to the situation of isolated cobbles colliding with a bedrock river bottom, but we expect that Phase I abrasion can be extended to the case of numerous cobbles colliding with each other. Identification of two-phase abrasion serves to organize existing field and laboratory data. It is clear that abrasion should contribute substantially to pebble mass loss and the production of fine sediment in a river (Phase I), even if it may be subordinate to size-selective sorting in driving downstream decreases in pebble diameter (Phase II). Two-phase abrasion resolves the shape-size paradox. Explicit accounting of shape in future abrasion studies will allow for a better understanding of the contributions of other factors. Future work could combine shape evolution with a mechanical abrasion model that considers collision energy and material properties [37,38], and explicitly model multi-particle collisions, to assess the robustness of our reported results. Constraining the kinematics of the abrasion process (coefficient $g$) might also allow one to infer a pebble's age from its shape by forward modeling of **Eq. 1** – if one can reasonably assume an initial condition – similar to morphometric dating commonly applied to hillslope scarps using the diffusion equation [39]. Such work could then serve to provide more quantitative bounds on past stream flows and climate conditions associated with river deposits on Earth and other planets. The recent discovery of rounded pebbles – apparently in Phase II of their shape evolution – in a rock outcrop on Mars, for example, was used to infer that an ancient river had abraded sediments during transport over kilometers [40,41]. More generally, the curvature-driven flow model connects the shape evolution of pebbles to a much broader class of problems governed by surface diffusion, such as the Khardar-Parisi-Zhang equation for surface growth [42]. This mathematical connection may be exploited to model pebble shape evolution under a range of boundary and initial conditions, by making use of existing numerical and analytical solutions.

**Supporting Information**
Supporting Information File S1 is an Excel spreadsheet file that contains all data on mass and shape evolution, from drum experiments and numerical simulations, that were used to generate Figure 4.

## Materials and methods
### Laboratory experiments
We performed four laboratory experiments in which single cuboids composed of oolitic limestone (well sorted calcarenite, for detailed lithological description see [43]), were abraded in a 1-m diameter, rotating "Los Angeles" drum. At specified rotation intervals (0,25,50…500, 550, 600) we imaged the particles from three orthogonal directions, and measured the corresponding dimensions (principal axes) and mass. Faces were marked with ink for repeated identification. Volume was computed from mass assuming homogeneous sample density. High sample porosity led to rapid erosion, such that a cuboid evolved to a sphere in approximately one day.

### Level set method
Here we simulated surface evolution (**Eq. 2**) under the linear combination of the Gaussian and the mean curvature flows, using the classical *level set method* [44]. Simulations used the Matlab Toolbox for Level Set Methods by Mitchell [27, 28]. Note that curvature-dependent flows require the computation of the second gradients of the surface; thus a polyhedron as an initial shape represents a singularity. In our numerical investigations we found that a superellipsoid with n < 20 is needed to start the computation.

### Chopping model
The *chopping model* is a discrete, stochastic algorithm where a sequence of local collisional events leads to shape evolution; it was originally presented by [22], where details can be found. For an alternative planar model see [45]. A particle is represented by the polyhedron $P$ and abrasion results due to discrete collisions by another polyhedron $P^*$. For each collision a volume $\Delta$ with lognormal distribution LogNorm($\Delta_0,\sigma_1$), mean value $\Delta_0$ (depending on the volume of $P$), and variation $\sigma_1$ is removed by intersecting $P$ with a plane having random orientation. Numerically, this step is carried out by a polyhedron splitting code embedded in an inverse iteration aimed to recover the prescribed volume $\Delta$. We remark that in the current version of our code $\Delta$ does not depend on the edge angles and thus the abraded volume is over-estimated for small edge angles. This is manifested in the differences on Figure 4c showing the evolution of the surface convexity index β; initially, with small edge angles close to 90 degrees, the computation predicts faster-than-realistic abrasion. The three types of collision events, shown in Fig. 1b, are modeled assuming that the directional distribution of collisions is uniform, in accordance with the geometric assumptions underlying the averaged PDE **Eq. 2**. (A) The impactor $P^*$ is large and flat; collision occurs between a face of $P^*$ and a vertex of $P$. Impact location on $P$ is selected randomly based on solid angles of the surface normal (the integrated Gaussian curvature). In this case sharp vertices are selected with high probability, and a vertex of $P$ is chopped off and replaced by a small face, normal to the randomly selected direction. (B) The impactor $P^*$ is large and thin; collision occurs between an edge of $P^*$ and an edge of $P$. Impact location on $P$ is selected randomly based on total product of edge length and edge angle (the integrated mean curvature). In this case sharp and long edges are selected with high probability, and an edge of $S$ is chopped off and replaced by a small, thin face, the normal of which is chosen uniformly in the range defined by the normals of the adjacent faces. (C) The impactor $P^*$ is much smaller than the object; collision occurs between a vertex of $P^*$ and a face of $P$. Impact location on $P$ is selected randomly based on surface area. For this case large faces are selected with high probability, and the selected face of $P$ retreats parallel to itself. This component is not relevant for the experiments examined here but is included for completeness. The average abrasion rate over many collisions is given by **Eq. 3**.

## Acknowledgements
We sincerely thank Mikael Attal for insightful comments that improved this manuscript. We gratefully acknowledge support for this work through OTKA grant T104601 to GD and AS, and NSF agreement EAR-1224943 to DJJ. The authors are indebted to Sarolta Bodor, Gyula Emszt, Bálint Pálinkás and Ottó Sebestyén for their valuable help with the experiments.

## References
1. Krynine P-D. On the Antiquity of "Sedimentation" and Hydrology. *Bulletin of the Geological Society America* 71:1721-1726 (1960) .
2. Parker G. Selective Sorting and Abrasion of River Gravel. 1. Theory. *J. Hydraul. Eng. - ASCE* 117(2):131–149 (1991).
3. Attal M, J. Lave. Pebble abrasion during fluvial transport: experimental results and implications for the evolution of the sediment load along rivers. J*ournal of Geophysical Research* 114:F04023. doi:10.1029/2009jf001328 (2009).
4. Krumbein W-C. The effects of abrasion on the size, shape and roundness of rock fragments. *The Journal of Geology* 49(5):482-520 (1941).

5. Lewin J, Brewer P. Laboratory simulation of clast abrasion. *Earth Surface Processes and Landforms* 27(2):145–164. doi:10.1002/esp.306 (2002).
6. Jerolmack D, Brzinski T. Equivalence of abrupt grain-size transitions in alluvial rivers and eolian sand seas: A hypothesis. *Geology* 38(8):719–722, doi:10.1130/G30922.1 (2010).
7. Boggs Jr. S. *Sedimentology and Stratigraphy* (Pearson Education: 2006).
8. Sternberg H. Untersuchungen Uber Langen-und Querprofil geschiebefuhrender Flusse. *Zeitschrift für Bauwesen* XXV:483-506 (1875).
9. Wentworth C-K. A Laboratory and Field Study of Cobble Abrasion. *The Journal of Geology* 27(7):507–521. doi:10.2307/30058414 (1919).
10. Kuenen P-H. Experimental Abrasion of Pebbles: 2. Rolling by Current. *The Journal of Geology* 64(4):336–368. doi:10.2307/30056065 (1956).
11. Kodama Y. Experimental-study of abrasion and its role in producing downstream fining in gravel-bed rivers. *Journal of Sedimentary Research Section A-Sedimentary Petrology* 64(1):76–85 (1994).
12. Ferguson R, Hoey T, Wathen S, Werritty A. Field evidence for rapid downstream fining of river gravels through selective transport. *Geology* 24(2):179–182 (1996).
13. Morris P, Williams D. A worldwide correlation for exponential bed particle size variation in subaerial aqueous flows. *Earth Surface Processes and Landforms* 24(9):835–847 (1999).
14. Hoey T, Bluck B. Identifying the controls over downstream fining of river gravels. *Journal of Sedimentary Research* 69(1):40–50 (1999).
15. Paola C, Parker G, Seal R, Sinha S-K, Southard J-B, Wilcock P-R. Downstream Fining by Selective Deposition in a Laboratory Flume. *Science* 258(5089):1757–1760 (1992).
16. Seal R, Paola C, Parker G, Southard J-B, Wilcock P-R. Experiments on downstream fining of gravel. 1. Narrow-channel runs. *J. Hydraul. Eng.-ASCE* 123(10):874–884 (1997).
17. Jerolmack D-J, Reitz M-D, Martin R-L. Sorting out abrasion in a gypsum dune field. *J. Geophys. Res.* 116(F2). doi:10.1029/2010JF001821 (2011).
18. Durian D, Bideaud H, Duringer P, Schröder A, Thalmann F, Marques C. What Is in a Pebble Shape? *Phys. Rev. Lett.* 97(2). doi:10.1103/PhysRevLett.97.028001 (2006).
19. Roth, A. E., C. M. Marques, and D. J. Durian. Abrasion of flat rotating shapes, *Phys. Rev. E* 83(3):031303. doi:10.1103/PhysRevE.83.031303 (2011).
20. Firey W-J. The shape of worn stones. *Mathematika* 21:1-11 (1974).
21. Bloore F.-J. The shape of pebbles. *Mathematical Geology* 9(2):113–122. doi:10.1007/BF02312507 (1977).
22. Domokos G, Sipos A-A, Várkonyi P. Continuous and discrete models for abrasion processes. *Periodica Polytechnica – Architecture* 40(1):3-8. (2009)
23. Blott, S.J., Pye, K. Particle shape: a review and new methods of characterization and classification Sedimentology 55:31-63. doi: 10.1111/j.1365-3091.2007.00892.x (2008).
24. Zingg T. Beitrag zur Schotteranalyse. *Schweiz. Mineral. Petrogr. Mitt* 15:39-140 (1935).
25. Wadell H. Volume, Shape and Roundness of Quartz Particles. *Journal of Geology* 43(3):250–280 (1935).
26. Várkonyi P, Domokos G. A general model for collision-based abrasion processes. *IMA J. Appl. Math.* 76 (1):47-56 (2011).
27. Mitchell I-M, Templeton J-A. A Toolbox of Hamilton-Jacobi Solvers for Analysis of Nondeterministic Continuous and Hybrid System. HSCC'05 Proceedings of the 8th international conference on Hybrid Systems: computation and control 480-494 (2005).
28. Mitchell I-M. The Flexible, Extensible and Efficient Toolbox of Level Set Methods. Journal of Scientific Computing 35:300-329 (2008).
29. Domokos G, Gibbons G-W. The evolution of pebble size and shape in space and time. Proc. R. Soc. A-Math. Phys. Eng. Sci. 468(2146):3059–3079. doi:10.1098/rspa.2011.0562 (2012).
30. Domokos G, Sipos A-A, Szabó Gy, Várkonyi P. Formation of sharp edges and planar areas

of asteroids by polyhedral abrasion. Astrophysical Journal 699(1):L13-L16 (2009).
31. Fourier J. La théorie analitique de la chaleux. Paris (1822).
32. John F. Partial differential equations. 4th ed (Springer: 1986).
33. Hamilton, R.S. Worn stones with flat sides, in: A tribute to Ilya Bakelman (College. Station, TX 1993), *Discourses Math. Appl*. **3** (1994), pp. 69–78.
34. Adams J. Wear of Unsound Pebbles in River Headwaters. *Science* 203(4376):171–172. doi:10.2307/1747672 (1979).
35. Szabó T., Fityus S., Domokos G. Abrasion model of downstream changes in grain shape and size along the Williams River, Australia. J.Geophyiscal Research/Earth Surface , DOI: 10.1002/jgrf.20142
36. Domokos,G., Gibbons, G.W. Geometrical and physical models of abrasion. ArXiv preprint: http://arxiv.org/abs/1307.5633 (2013).
37. Anderson R. Erosion profiles due to particles entrained by wind - application of an eolian sediment-transport model. *Geological Society of America Bulletin* 97(10):1270–1278 (1986).
38. Wang Z-T, Wang H-T, Niu Q-H, Dong Z-B, Wang T. Abrasion of yardangs. *Phys. Rev. E*. 84(3). doi:10.1103/PhysRevE.84.031304 (2011) .
39. Colman S-M, Watson K. Ages Estimated from a Diffusion Equation Model for Scarp Degradation. *Science* 221(4607):263–265. doi:10.1126/science.221.4607.263 (1983).
40. Williams R-M-E. et al. Martian Fluvial Conglomerates at Gale Crater. *Science* 340(6136):1068–1072. doi:10.1126/science.1237317 (2013) .
41. Jerolmack D-J. Pebbles on Mars. *Science* 340(6136):1055–1056. doi:10.1126/science.1239343 (2013).
42. Kardar M, Parisi G, Zhang Y-C. Dynamic scaling of growing interfaces. *Phys. Rev. Lett*. 56:889-892 (1986).
43. Török Á. Oolitic limestone in polluted atmospheric environment in Budapest: weathering phenomena and alterations in physical properties. *Geological Society London Special Publications* 205: 363-379 (2002).
44. Giga Y. *Surface Evolution Equations, a Level Set Approach* (Birkhäuser Verlag: 2006).
45. Krapivsky P.L., Redner S. Smoothing rock by chipping. *Physical Review E.* Vol 75(3 Pt 1):031119 DOI:10.1103/PhysRevE.75.031119